# Intrinsic Dielectric Response in Ferroelectric Nano-Crystals

M. M. Saad[1], P. Baxter[1], R. M. Bowman[1], J. M. Gregg[1]

F. Morrison[2] and J. F. Scott[2]

[1]Department of Pure & Applied Physics, Queen's University Belfast, Belfast BT7 1NN, UK

[2]Symetrix Centre for Ferroics, Department of Earth Science, University of Cambridge, Cambridge CB2 3EQ, UK

ABSTRACT

Measurements on 'free-standing' single-crystal barium titanate capacitors with thicknesses down to 75 nm show a dielectric response typical of large single crystals, rather than conventional thin films. There is a notable absence of any broadening or temperature shift of the dielectric peak or loss tangent. Peak dielectric constants of ca. 25,000 are obtained, and Curie-Weiss analysis demonstrates 1st order transformation behavior. This is in surprising contrast to results on conventionally deposited thin film capacitor heterostructures, which show large dielectric peak broadening and temperature shifts [*e.g.* C. B. Parker, J.-P. Maria, and A. I. Kingon, Appl. Phys. Lett. **81**, 340 (2002)], as well as an apparent change in the nature of the paraelectric-ferroelectric transition from 1st to 2nd order. Our data are compatible with the recent model by Bratkovsky and Levanyuk, which attributes dielectric peak broadening to gradient terms that will exist in any thin film capacitor heterostructure, either through defect profiles introduced during growth, or through subtle asymmetry between top and bottom electrodes. The observed recovery of 1st order transformation behavior is consistent with the absence of significant substrate clamping in our experiment, as modeled by Pertsev *et al.* Phys. Rev. Lett., **80**, 1988 (1998), and illustrates that the 2nd order behavior seen in conventionally deposited thin films cannot be attributed to the effects of reduced dimensionality in the system, nor to the influence of an intrinsic universal interfacial capacitance associated with the electrode-ferroelectric interface.

Observations on ferroelectric thin films always show that a significant broadening of the Curie anomaly (the peak in dielectric constant with respect to temperature) occurs as films are made thinner [1-4]. Typically, parameters expressing peak 'diffuseness' are many orders of magnitude greater for films ~100 nm or thinner, than for bulk [2,3]. In addition, the broadening of the dielectric peak is associated with an apparent change in the nature of the paraelectric-ferroelectric phase transition from $1^{st}$ to $2^{nd}$ order [4-6]. Until now, the origins of these changes in functional behavior have not been clear, but several pertinent ideas exist in literature:

(i) It is known [7-10] that two-dimensional effects broaden out any first-order discontinuities at phase transition temperatures and make systems appear second-order. While thin films with thickness ~100nm are clearly not two-dimensional, ferroelectrics are dominated by long-range Coulombic forces, and it is difficult to be certain about the thickness at which effects due to the reduction of system dimensionality may become evident; while some understanding of the thickness limits on the existence of static properties such as polarization and ferroelectricity have recently been established [6, 11], this has not been extended to an understanding of the thickness at which dimensionality-induced changes in intrinsic functional *dynamics* might be observed.

(ii) Since the earliest work on thin film ferroelectrics, the influence of a parasitic 'interfacial capacitance' has been identified [12]. This manifests itself as a small capacitive component that acts electrically in series with the rest of the film, and as a result suppresses the overall observed dielectric constant in the system as a whole. Moreover, since series addition of dielectric response follows $1/C_T = \sum_i 1/C_i$, the influence of the interfacial capacitance is greatest at the Curie anomaly, and less influential far from the anomaly, automatically generating a peak smearing effect. The potential origins of the interfacial capacitance fall into two broad classes – those related to the inherent and unavoidable physics of the ferroelectric-electrode boundary [13-22] and those which may be induced through specific processing issues in the growth of the heteroepitaxial system [23-28] (for example a chemically or microstructurally distinct interfacial region, or the influence of grain boundaries).

(iii) The strain state of thin film ferroelectrics can be altered by the nature of the lower electrode, or substrate material onto which it is grown. Landau-Ginzburg-Devonshire considerations, assuming homogeneous strain alone [29, 30], have shown that this can also result in the paraelectric-ferroelectric phase transition changing from $1^{st}$ to $2^{nd}$ order. In addition, there is a growing body of experimental evidence to suggest that minimizing the substrate / electrode mismatch strain, or reducing substrate clamping effects can recover 'bulk-like' properties to some degree [31, 32];

(iv) It has been widely recognized that the presence of a surface in an otherwise centric material is formally equivalent to a field; for example, in yttrium barium copper oxide (YBCO) treating the surface as polar mm2 point group symmetry, with field normal to the surface, rather than the mmm centric bulk symmetry [33], permitted the theoretical analysis of unexplained photovoltaic data [34, 35]. Very recently, this symmetry argument has been developed into a detailed theory for ferroelectric thin films by Bratkovsky and Levanyuk [36]:

Assuming an arbitrary gradient in a scalar quantity across the film (such as defect concentration, chemical concentration, density or temperature) they show that an "extra" field-like term exists in the free-energy expression:

$$f_c = -\gamma P \nabla c \qquad (1)$$

leading to an equation of state:

$$A(z)P + BP^3 - g\frac{d^2P}{dz^2} - D\nabla_\perp^2 P = E_0 + 4\pi(\overline{P} - P) + \gamma\frac{dc}{dz} \qquad (2)$$

and they calculate numerically, for plausible parameters, the strong broadening effect this has on the dielectric response near the Curie temperature ($T_C$); a rather complete washing-out of the dielectric peak is predicted. We note that the disorder models of Harris [37] and Stinchcombe [38] may also be relevant in the context of such broadening of phase transitions (second-order or tricritical).

The Bratkovsky and Levanyuk model is particularly timely, for although phenomenological treatments involving gradient terms have been investigated before [39, 40], recent experiments by Ma and Cross [41-43] have highlighted the likelihood that strain gradient terms (flexoelectric terms) could dominate ferroelectric thin film functional response – a view subsequently reinforced by Catalan *et al.* [44], who predicted that gradient terms that were associated with progressive relaxation of coherent interfacial strains, should smear out dielectric response around $T_C$.

Gradients in oxygen vacancies ($\nabla c$) could also play a significant role. Reference to Scott and Dawber [45], suggests that in materials such as barium titanate, a bulk diffusion-limited depth dependence over the outermost 10nm with a peak value of 8 x $10^{21}$ cm$^{-3}$, and a '$c = \exp(z^{-6/5})$' grain-boundary limited dependence, at depths 10 - 30 nm or deeper, exist near to ferroelectric surfaces. Thus, one could fit the Bratkovsky-Levanyuk model with two separate step-like dependences of $\nabla c$. However, it may be that vacancy concentration gradients simply represent a special case of strain gradients, since vacancies produce significant inhomogeneous strain. Recent work by Balzar *et al.* [46] supports this view.

In the present study, by examining the functional properties of single crystal lamellae cut using a focused ion-beam microscope (FIB), we largely eliminate both interfacial strain and scalar concentration gradients associated with conventional thin film deposition. The results show that, in heteroepitaxial systems investigated to date, dielectric broadening and apparent 2$^{nd}$ order phase transition behavior must be due to these extrinsic factors; they are not a result of either the effect of reduced dimensionality, or the influence of an unavoidable intrinsic interfacial capacitance associated with the electrode-ferroelectric interface.

Commercially obtained polished single crystal square plates of BaTiO$_3$ were used as starting materials. These crystals were placed in a FEI200TEM FIB, such that the incident primary gallium (Ga)-beam was parallel to the top polished surface, and perpendicular to one of the sides of the plates. A series of milling steps was then performed, to create a thin lamella parallel to the top polished surface, and connected to the rest of the BaTiO$_3$ single crystal along three of its four sides. A series of such lamellae of varying thickness was produced from the same single crystal, before furnace-annealing in air at 700$^o$C for 1 hour. Gold was then evaporated onto the crystal from two directions both at 45$^o$ to the polished crystal surface, such that both sides of the dielectric lamella were fully coated. The coated crystal was once again placed in the FIB, but with the polished crystal surface perpendicular to the Ga-beam. Rectangular electrodes (5μm x 7μm) on the top surface of the lamellae were isolated by milling through to the dielectric; further milling defined gold strips that connected the lamellae electrodes to contact pads. For testing, a micromanipulator was then used to make contact to the pads, with the other contact made by wire-bonding to the rest of the gold (electrically shorted to the back face of the parallel-plate capacitors). More details of the fabrication procedure adopted, and a demonstration that the thermal anneal fully recovered any damage caused to the lamellae through Ga-ion implantation can be found elsewhere [47]. Capacitance and loss tangent were measured using a Hewlett-Packard HP4284A LCR meter and HP4192A impedance analyser with applied test voltages of 100mV, with temperature variable measurements made using a programmable hotplate (heating rate ~ 1 Kmin$^{-1}$). For measurement of temperature, a K-type thermocouple was mounted on the surface of a dummy $BaTiO_3$ crystal of identical dimensions to that containing the test capacitors.

Figure 1 illustrates the frequency dependence of the measured capacitance from a thin single crystal lamella, at several temperatures. In all cases, the spectral profile is reasonably flat, with no indication of space charge activity. The figure also illustrates the large variation in measured capacitance with temperature. This is further elaborated upon in figure 2, where the dielectric constant and loss behavior from the thinnest of the lamellae examined (75nm) is plotted at 10kHz against temperature. As can be seen, the dielectric constant goes through a distinct sharp anomaly at the Curie temperature, mirroring the response that is typically seen in bulk single crystal material. The temperature of the Curie anomaly was found to be ~395 ± 5 K for all the lamellae examined (between ~ 450nm and ~ 75nm in thickness). Such observations are completely different from any published data on conventionally grown heteroepitaxial capacitor structures across the same thickness range. Also worthy of note are the absolute magnitudes of the dielectric constant, and the low loss tangents, again more reminiscent of bulk single crystal behavior than conventional thin film. Figure 3 is a Curie-Weiss plot for the dielectric constant data above the dielectric peak, illustrating that the thin lamellae adhere well to Curie-Weiss behavior, with $T_0$ below the apparent $T_C$ – a sufficient condition for the description of 1$^{st}$ order transformation behavior [48].

Overall the dielectric response (in terms of absolute magnitude of dielectric constant, variation with temperature and with frequency, apparent 1$^{st}$ order transformation behavior) is extremely similar to that expected from bulk single crystal, with none of the changes in behavior seen in conventionally grown systems, universally accepted as 'size effects'. While this is striking, recent work on carefully grown thin film systems have hinted towards the result: Dittman *et al.* [31] have shown that completely epitaxial growth of both the thin film ferroelectric and ceramic electrodes can significantly sharpen the form of the peak in dielectric constant against temperature, and raise the magnitude of the peak dielectric constant closer to bulk / single crystal values than had previously been thought possible. Takashima *et al.* made similar observations in SrTiO$_3$ thin films [32]. No one, however, has shown that full recovery of bulk-like functional behavior can occur for ferroelectrics in the sub-100nm thickness regime, as has been illustrated here. This result therefore acts as a paradigm for those involved in the thin film ferroelectric growth community, and for those seeking to utilise ferroelectric thin films in microelectronic applications.

In conclusion, the focused ion beam microscope has been used to fabricate thin lamellae of single crystal BaTiO$_3$, with gold evaporated on the lamellar walls to form parallel-plate capacitor structures. The functional characterisation of these capacitors demonstrated that the smearing of the peak in dielectric constant, and the apparent change in the nature of the phase transition from $1^{st}$ to $2^{nd}$ order are neither related to the fundamentals of reduced dimensionality nor to any unavoidable intrinsic interfacial capacitance associated with the electrode-ferroelectric interface. Rather, the influence of homogeneous strain through strain coupling to a substrate, or gradient terms either associated with chemical, defect or strain gradients are the primary suspects.

The authors acknowledge the Engineering and Physical Sciences Research Council (EPSRC) and the International Research Centre for Experimental Physics (IRCEP) for financial support.

**Figure Captions:**

**Figure 1:** Frequency-dependence of the capacitance measured at 3 different temperatures for thin $BaTiO_3$ lamellae. All lamellae investigated showed the distinct lack of spectral dependence illustrated here. Strong temperature-dependence is, however, evident.

**Figure 2:** The calculated relative permittivity (a) and measured loss tangent (b) as a function of temperature at 10kHz for a 75nm-thick single crystal $BaTiO_3$ capacitor. The absolute magnitude of the dielectric constant, its demonstration of a sharp Curie anomaly, and the occurrence of the anomaly around 395K demonstrate that the thin lamella behaves like a bulk single crystal, and not as a conventionally grown thin film.

**Figure 3:** Above the observed Curie anomaly, the dielectric response of the thin lamellae showed Curie-Weiss response, with $T_0$ below $T_C$.

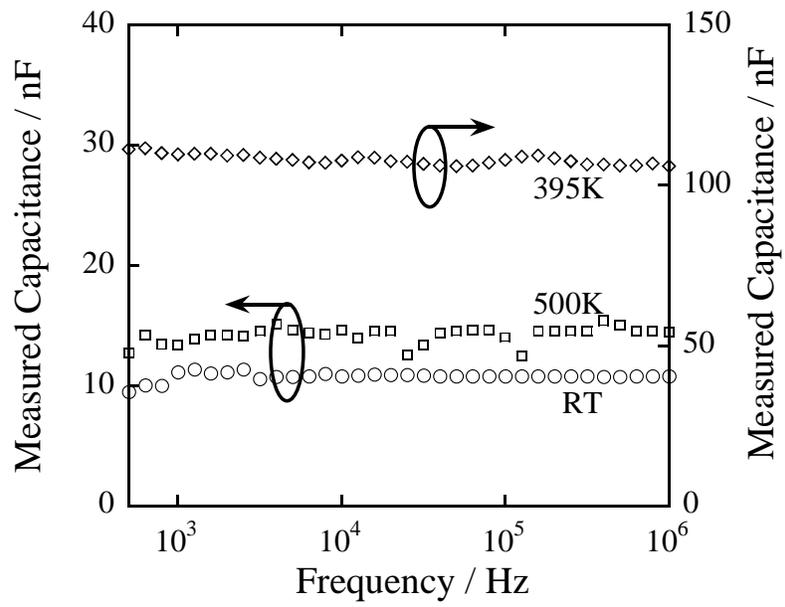

Figure 1 of 3: Saad *et al.* submitted to PRL

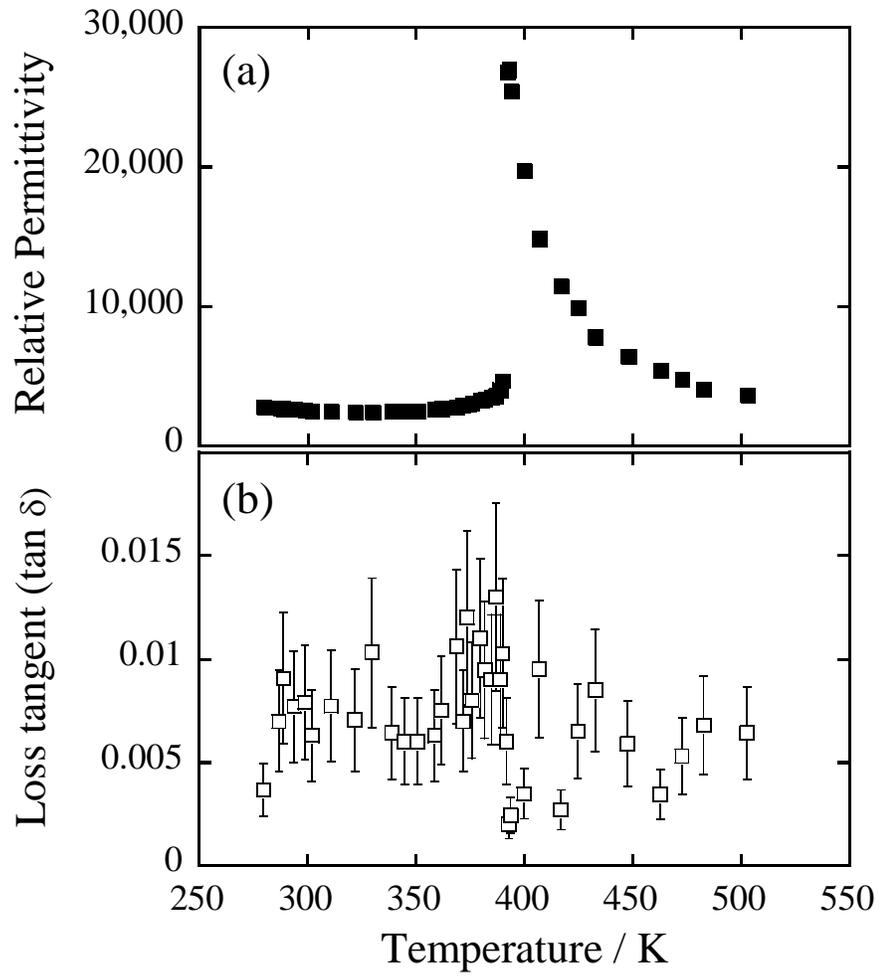

Figure 2 of 3: Saad *et al.* submitted to PRL

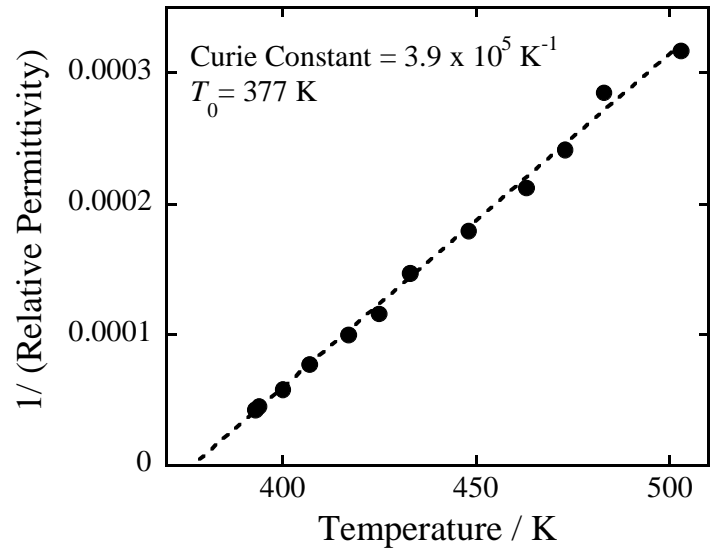

Figure 3 of 3: Saad *et al.* submitted to PRL